# Guided Data Repair [*]

Mohamed Yakout [1,2]   Ahmed K. Elmagarmid [2]   Jennifer Neville [1]   Mourad Ouzzani [1]   Ihab F. Ilyas [2,3]

[1]Purdue University   [2]Qatar Computing Research Institute, Qatar Foundation   [3]University of Waterloo

[1]{myakout, neville, mourad}@cs.purdue.edu   [2]aelmagarmid@qf.org.qa   [3]ilyas@uwaterloo.ca

## ABSTRACT

In this paper we present GDR, a Guided Data Repair framework that incorporates user feedback in the cleaning process to enhance and accelerate existing automatic repair techniques while minimizing user involvement. GDR consults the user on the updates that are most likely to be beneficial in improving data quality. GDR also uses machine learning methods to identify and apply the correct updates directly to the database without the actual involvement of the user on these specific updates. To rank potential updates for consultation by the user, we first group these repairs and quantify the utility of each group using the decision-theory concept of value of information (VOI). We then apply active learning to order updates within a group based on their ability to improve the learned model. User feedback is used to repair the database and to adaptively refine the training set for the model. We empirically evaluate GDR on a real-world dataset and show significant improvement in data quality using our user guided repairing process. We also, assess the trade-off between the user efforts and the resulting data quality.

## 1. INTRODUCTION

Poor data quality is a fact of life for most organizations and can have serious implications on their effectiveness [1]. An example critical application domain is healthcare, where incorrect information about patients in an Electronic Health Record (EHR) may lead to wrong treatments and prescriptions that may cause severe medical problems.

A recent approach for repairing dirty databases is to use data quality rules in the form of database constraints to identify tuples with errors and inconsistencies and then use these rules to derive updates to these tuples. Most of the existing data repair approaches (e.g., [2, 6, 7, 16]) focus on providing fully automated solutions using different heuristics to select updates that would introduce minimal changes to the data, which could be risky especially for critical data. To guarantee that the best desired quality updates are applied to the database, users (domain experts) should be involved to confirm updates. This highlights the increasing

[*]This research was supported by QCRI, and by NSF Grant Numbers IIS 0916614 and IIS 0811954, and by the Purdue Cyber Center.

| Name | SRC | STR | CT | STT | ZIP |
|---|---|---|---|---|---|
| t1: Jim | H1 | REDWOOD DR | MICHIGAN CITY | MI | 46360 |
| t2: Tom | H2 | REDWOOD DR | WESTVILLE | IN | 46360 |
| t3: Jeff | H2 | BIRCH PARKWAY | WESTVILLE | IN | 46360 |
| t4: Rick | H2 | BIRCH PARKWAY | WESTVILLE | IN | 46360 |
| t5: Joe | H1 | BELL AVENUE | FORT WAYNE | IN | 46391 |
| t6: Mark | H1 | BELL AVENUE | FORT WAYNE | IN | 46825 |
| t7: Cady | H2 | BELL AVENUE | FORT WAYNE | IN | 46825 |
| t8: Sindy | H2 | SHERDEN RD | FT WAYNE | IN | 46774 |

(a) Data

$\phi_1 : (\text{ZIP} \rightarrow \text{CT}, \text{STT}, \{46360 \parallel \text{MichiganCity}, \text{IN}\})$
$\phi_2 : (\text{ZIP} \rightarrow \text{CT}, \text{STT}, \{46774 \parallel \text{NewHaven}, \text{IN}\})$
$\phi_3 : (\text{ZIP} \rightarrow \text{CT}, \text{STT}, \{46825 \parallel \text{FortWayne}, \text{IN}\})$
$\phi_4 : (\text{ZIP} \rightarrow \text{CT}, \text{STT}, \{46391 \parallel \text{Westville}, \text{IN}\})$
$\phi_5 : (\text{STR}, \text{CT} \rightarrow \text{ZIP}, \{\_, \text{FortWayne} \parallel \_\})$

(b) CFD Rules

**Figure 1: Example data and rules**

need for a framework that combines the best of both worlds. The framework will automatically suggest updates while efficiently involve users to guide the cleaning process.

### 1.1 Motivation

Consider the following example. Let Relation Customer(Name, SRC, STR, CT, STT, ZIP) specifies personal address information Street (STR), City (CT), State (STT) and (ZIP), in addition to the source (SRC) of the data or the data entry operator. An instance of this relation is shown in Figure 1.

Data quality rules can be defined in the form of Conditional Functional Dependencies (CFDs) as described in Figure 1(b). A CFD is a pair consisting of a standard Functional Dependency (FD) and a pattern tableau that specifies the applicability of the FD on parts of the data. For example, $\phi_1 - \phi_4$ state that the FD ZIP $\rightarrow$ CT, STT (i.e., zip codes uniquely identify city and state) holds in the *context* where the ZIP is 46360, 46774, 46825 or 46391. Moreover, the pattern tableau enforces bindings between the attribute values, e.g., if ZIP= 46360, then CT= 'Michigan City'. $\phi_5$ states that the FD STR, CT $\rightarrow$ ZIP holds in the context where CT = 'Fort Wayne', i.e., street names uniquely identify the zip codes whenever the city is 'Fort Wayne'. Note that all the tuples in Figure 1 have violations.

Typically, a repairing algorithm will use the rules and the current database instance to find the best possible repair operations or updates. For example, $t_5$ violates $\phi_4$ and a possible update would be to either replace CT by 'Westville' or replace ZIP by 46825, which would make $t_5$ fall in the context of $\phi_3$ and $\phi_5$ but without violations. To decide which update to apply, different heuristics can be used [2, 16].

However, automatic changes to data can be risky espe-cially if the data is critical, e.g., choosing the wrong value among the possible updates. On the other hand, involving the user can be very expensive because of the large number of possibilities to be verified. Since automated methods for data repair produce far more updates than one can expect the user to handle, techniques for selecting the most useful updates for presentation to the user become very important.

Moreover, to efficiently involve the user in guiding the cleaning process, it is helpful if the suggested updates are presented in groups that share some contextual information. This will make it easier for the user to provide feedback. For example, the user can quickly inspect a group of tuples where the value 'Michigan City' is suggested for the CT attribute. Similar grouping ideas have been explored in [19].

In the example in Figure 1, let us assume that a cleaning algorithm suggested two groups of updates. In the first group, the updates suggest replacing the attribute CT with the value 'Michigan City' for $t_2, t_3$, and $t_4$ while in the second group they suggest replacing the attribute ZIP with the value 46825 for $t_5$ and $t_8$. Let us assume further that we were able to obtain the user feedback on the correct values for these tuples; namely that the user has confirmed 'Michigan City' as a correct value of CT for $t_2, t_3$, but as incorrect for $t_4$, and 46825 as the correct value of ZIP for $t_5$, but as incorrect for $t_8$. In this case, consulting the user on the first group, which has more correct updates, is better and would allow for faster convergence to a cleaner database instance as desired by the user. The second group will not lead for such fast convergence.

Finally in our example, we could recognize correlations between the attribute values in a tuple and the correct updates. For example, when SRC = 'H2', the CT attribute is incorrect most of the time, while the ZIP attribute is correct. This is an example of recurrent mistakes that exist in real data. Patterns like that with correlations between the original tuple and the correct updates, if captured by a machine learning algorithm, can reduce user involvement.

The key challenge in involving users is to determine *how* and in *what* order suggested updates should be presented to them. This requires developing a set of principled measures to estimate the improvement in quality to reason about the selection process of possible updates as well as investigating machine learning techniques to minimize user effort. The goal is to achieve a good trade-off between high quality data and minimal user involvement.

In this paper, we propose to tackle the problem of data cleaning from a more realistic and pragmatic viewpoint. We present GDR, a framework for guided data repair, that interactively involves the user in guiding the cleaning process alongside existing automatic cleaning techniques. The goal is to effectively involve users in a way to achieve better data quality as quickly as possible. The basic intuition is to continuously consult the user for updates that are most beneficial in improving the data quality as we go.

We use CFDs [3] as the data quality rules to derive candidate updates. CFDs have proved to be very useful for data quality and triggered several efforts e.g., [9, 13], for their automatic discovery as well as making them a practical choice for data repair techniques.

## 1.2 Problem Definition

We consider a database instance $D$ with a relational schema $S$. Each relation $R \in S$ is defined over a set of attributes $attr(R)$ and the domain of an attribute $A \in attr(R)$ is denoted by $dom(A)$. We also consider a set of data quality rules $\Sigma$ that represent data integrity semantics. In this paper, we consider rules in the form of CFDs.

A CFD $\phi$ over $R$ can be represented by $\phi : (X \rightarrow Y, t_p)$, where $\{X \cup Y\} \subseteq attr(R)$, $X \rightarrow Y$ is a standard FD and $t_p$ is a tuple pattern for all attributes in $X$ and $Y$. For each $A \in (X \cup Y)$, the value of the attribute $A$ for the tuple pattern $t_p$, $t_p[A]$, is either a constant 'a' $\in dom(A)$, or '−' representing a variable value. We denote $X$ as $LHS(\phi)$ (left hand side) and $Y$ as $RHS(\phi)$ (right hand side). Examples of CFD rules are provided in Figure 1.

We assume that CFDs are provided in the normal form [7]. If $\phi : (X \rightarrow Y, t_p)$ and $Y = \{A_1, A_2, \dots\}$, then the normal form is to split $\phi$ into $\phi_1 : (X \rightarrow A_1, t_p), \phi_2 : (X \rightarrow A_2, t_p), \dots$. A CFD $\phi : (X \rightarrow A, t_p)$ is said to be *constant*, if $t_p[A] \neq$ '−'. Otherwise, $\phi$ is a *variable* CFD. Constant rule can be violated by a single tuple, while variable ones (similar to FDs) are violated by multiple tuples. For example in Figure 1, $\phi_1$ is a constant CFD, while $\phi_5$ is a variable CFD.

We address the following problems:

- The use of the data quality rules $\Sigma$ to generate candidate updates for the tuples that are violating $\Sigma$. The rules can be either given or discovered by an automatic discovery technique (e.g., [9, 13]). Usually, the automatic discovery techniques employ thresholds on the confidence of the discovered rules. In this setting, the user is the one to guide the repairing process and we assume that user decisions are consistent with $\Sigma$.

- Deciding upon the best groups of updates—as mentioned in Section 1.1— to be presented to the user during an interactive process for faster convergence and higher data quality.

- Applying active learning to learn user feedback and use the learned models to decide upon the correctness of the suggested updates without user's involvement.

## 1.3 Summary of Contributions

We summarize our contributions as follows:

- We introduce GDR, a framework for data repair, that selectively acquire user feedback on suggested updates. User feedback is used to train the GDR machine learning component that can take over the task of deciding the correctness of these updates. (Section 2)

- We propose a novel ranking mechanism for suggested updates that applies a combination of decision theory and active learning in the context of data quality to reason about such task in a principled manner. (Section 4)

- We use the concept of value-of-information (VOI) [18] from decision theory to develop a mechanism to estimate the update benefit from consulting the user on a group of updates. We quantify the data quality loss by the degree of violations to the rules. The benefit of a group of updates can be then computed by the difference between the data quality loss before and after user feedback. Since we do not know the user feedback beforehand, we develop a set of approximations that allow efficient estimations. (Section 4.1)

- We apply active learning to order the updates within a group such that the updates that can strengthen the prediction capabilities of the learned model the most come first. To this end, we assign to each suggested update an uncertainty score that quantifies the benefit to the prediction model, learning benefit, when the update is labeled. (Section 4.2)

We conduct an extensive experimental evaluation on real datasets that shows the effectiveness of GDR in allowing fast convergence to a better quality database with minimal user intervention. (Section 5)

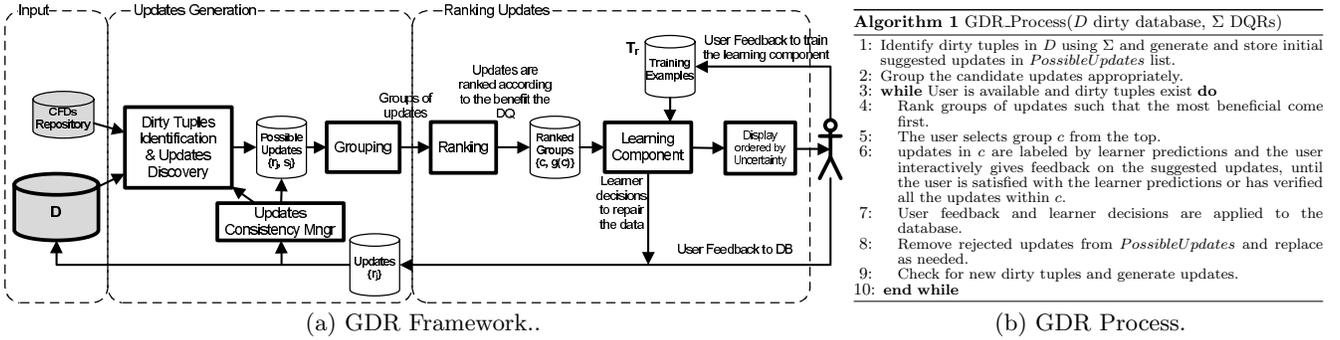

(a) GDR Framework..  (b) GDR Process.

Figure 2: GDR framework and process.

## 2. SOLUTION OVERVIEW

Figure 2(a) shows the GDR framework and the cleaning process is outlined in Procedure 2(b).

GDR *guides* the user to focus her efforts on providing feedback on the updates that would improve quality faster, while the user *guides* the system to automatically identify and apply updates on the data. This continuous feedback process, illustrated in steps 3-10 (Procedure 1), runs while there are dirty tuples and the user is available to give feedback.

In Step 1, all dirty tuples that violate the rules are identified and a repairing algorithm is used to generate candidate updates. In step 2, we group the updates for the user in a way that makes it easier for a batch inspection.

The interactive loop in steps 3-10 starts with ranking the groups of updates such that groups that are more likely to move the database to a cleaner state faster come first. The user will then pick one of the top groups ($c$) in the list and provide feedback through an interactive active learning session (step 6). (The ranking mechanism and active learning are discussed in Section 4.)

In step 7, all decisions on suggested updates, either made by the user or the learner, are applied to the database. In step 8, the list of candidate updates is modified by replacing rejected updates and generating new ones for emerging dirty tuples because of the applied updates.

After getting the user feedback, the violations are recomputed by the consistency manager and new updates may be proposed. The assumption is that if the user verifies all the database cells then the final database instance is consistent with the rules. This guarantees that we are always making progress toward the final consistent database and the process will terminate.

## 3. GENERATING CANDIDATE UPDATES

In this section, we outline the different steps involved in suggesting updates, maintaining their consistency when applied to the database, and grouping them for the user. More details of this section are given in Appendix A.

**Dirty tuples identification and updates discovery:** Once a set $\Sigma$ of CFDs is defined, *dirty* tuples can be identified through violations of $\Sigma$ and stored in a $DirtyTuples$ list. A tuple $t$ is considered dirty if $\exists \phi \in \Sigma$ such that $t \not\models \phi$, i.e., $t$ violates rule $\phi$.

We implemented an *on demand* update discovery process based on the mechanism described in [7] for resolving CFDs violations and generating candidate updates. This process is triggered to suggest an update for $t[A]$, the value of attribute $A$ in tuple $t$. Initially, the process is called for all dirty tuples and their attributes. Later during the interactions with the user, it is triggered by the consistency manager as a consequence of receiving user feedback.

The generated updates are tuples in the form $r_j = \langle t, A, v, s_j \rangle$ stored in the $PossibleUpdates$ list, where $v$ is the suggested value in $t[A]$ and $s_j$ is the *update score*. $s_j \in [0..1]$ is assigned to each update $r_j$ by an *update evaluation function* to reflect the certainty of the repairing technique about the suggested update. The evaluation function used in [2, 7] is the closeness in distance between the original and suggested values using some domain distance function.

**Updates Consistency Manager:** Once an update $r = \langle t, A, v, s \rangle$ is confirmed to be correct, either by the user or the learning component, it is immediately applied to the database resulting into a new database instance. Consequently, (i) new violations may arise and hence the on demand update discovery process needs to be triggered for the new dirty tuples, and (ii) some of the already suggested updates that are not verified yet may become inconsistent since they were generated according to a different database instance. For example, in Figure 1, two updates are proposed: $r_1$ replaces $t_6[\text{ZIP}] = 46391$ and $r_2$ replaces $t_6[\text{CT}] = $"FT Wayne". If a feedback is received confirming $r_1$, then $r_2$ is not consistent with the new database instance and the rules anymore since $t_6$ will fall in the context of $\phi_4$. The on demand process can then find a consistent update $r'_2$ that corresponds to replacing $t_6[\text{CT}]$ by 'Westville', and $r_2$ will be discarded in favor of $r'_2$.

Since GDR is meant for repairing online databases, the consistency manager will need to be informed (e.g., through database triggers) with any newly added or modified tuples so it can maintain the consistency of the suggested updates. In fact, GDR can be used in *monitoring* data entries and immediately suggesting updates during the data entry process. We do not discuss this issue further due to space limitation.

**Grouping Updates:** There are two reasons for the grouping: (i) Providing a *useful-looking* set of updates with some common contextual information will be easier for the user to handle and process. (ii) Providing a machine learning algorithm with a group of training examples that have some correlations due to the grouping will increase the prediction accuracy compared with just providing random, unrelated examples. Similar grouping ideas have been explored in [19]. We use a grouping function where the tuples with the same update value in a given attribute are grouped together.

## 4. RANKING AND DISPLAYING SUGGESTED UPDATES

In this section, we introduce the key concepts of GDR, namely the ranking and learning components (Figure 2(a)), which describe how GDR interacts with the user to get feedback on suggested updates. The task of these components is to devise how to best present the updates to the user, in a way that will provide the *most benefit* for improving the quality of the data. To this end, we apply the concept of value of information (VOI) [18] from decision theory, combined with an active learning approach, to choose a ranking in a principled manner.

## 4.1 VOI-based Ranking

At any iteration of the process outlined in Procedure 2(b), there will be several possible suggested updates to forward to the user. As discussed in the previous section, these updates are grouped into groups $\{c_1, c_2 \ldots\}$.

VOI is a mean of quantifying the potential benefit of determining the true value of some unknown. At the core of VOI is a *loss* (or utility) function that quantifies the desirability of a given level of database quality. To make a decision on which group to forward first to the user, we compare data quality loss *before* and after the user works on a group of updates. More specifically, we devise a *data quality* loss function, $L$, based on the quantified violations to the rules $\Sigma$. Since the exact loss in quality cannot be measured, as we do not know the correctness of the data, we develop a set of approximations that allow for efficient estimation of this quality loss. Before we proceed, we need first to introduce the notion of database violations.

DEFINITION 1. *Given a database $D$ and a CFD $\phi$, we define the tuple $t$ violation w.r.t $\phi$, denoted $vio(t, \{\phi\})$, as follows:*

$$vio(t, \{\phi\}) = \begin{cases} 1 & \text{, if } \phi \text{ is a constant CFD.} \\ \text{Number of tuples } t' \\ \text{that violate } \phi \text{ with } t & \text{, if } \phi \text{ is a variable CFD.} \end{cases}$$

*Consequently, the total violations for $D$ with respect to $\Sigma$ is:*

$$vio(D, \Sigma) = \sum_{\phi \in \Sigma} \sum_{t \in D} vio(t, \{\phi\}).$$

The definition for the variable CFDs is equivalent to the pairwise counting of violations discussed in [7]. The violation can be scaled further using a weight attached to the tuple denoting its importance for the business to be clean.

**Update Benefit**: Given a database instance $D$ and a group $c = \{r_1, \ldots, r_J\}$. If the system receives a feedback from the user on $r_j$, there are two possible cases: either the user confirms $r_j$ to be applied or not. We denote the two corresponding database instances as $D^{r_j}$ and $D^{\bar{r_j}}$, respectively. Assuming that the user will confirm $r_j$ with a probability $p_j$, then the expected data quality loss after consulting the user on $r_j$ can be expressed by: $p_j L(D^{r_j}) + (1 - p_j) L(D^{\bar{r_j}})$. If we further assume that all the updates within the group $c$ are independent then the update benefit $g$ (or data quality gain) of acquiring user feedback for the entire group $c$ can be expressed as:

$$g(c) = L(D|c) - \sum_{r_j \in c} [\, p_j \, L(D^{r_j}) + (1 - p_j) \, L(D^{\bar{r_j}}) \,] \quad (1)$$

where $L(D|c)$ is the current loss in data quality given that $c$ is suggested. To simplify our analysis, we assumed that these updates are independent. Taking into account these dependencies would require to model the full joint probabilities of the updates, which will lead to a formulation that is computationally infeasible due to the exponential number of possibilities.

**Data Quality Loss (L)**: We define quality loss as inversely proportional to the degree of satisfaction of the specified rules $\Sigma$. To compute $L(D|c)$, we first need to measure the *quality loss* with respect to $\phi \in \Sigma$, namely $ql(D|c, \phi)$. Assuming that $D^{opt}$ is the clean database instance desired by the user, we can express $ql$ by:

$$ql(D|c, \phi) = 1 - \frac{|D \models \phi|}{|D^{opt} \models \phi|} = \frac{|D^{opt} \models \phi| - |D \models \phi|}{|D^{opt} \models \phi|} \quad (2)$$

where $|D \models \phi|$ and $|D^{opt} \models \phi|$ are the numbers of tuples satisfying the rule $\phi$ in the current database instance $D$ and $D^{opt}$, respectively. Consequently, the data quality loss, given $c$, can be computed for Eq. 1 as follows:

$$L(D|c) = \sum_{\phi_i \in \Sigma} w_i \times ql(D|c, \phi_i). \quad (3)$$

where $w_i$ is a user defined weight for $\phi_i$. These weights are user defined parameters. In our experiments, we used the values $w_i = \frac{|D(\phi_i)|}{|D|}$, where $|D(\phi_i)|$ is the number of tuples that fall in the context of the rule $\phi_i$. The intuition is that the more tuples fall in the context of a rule, the more important it is to satisfy this rule. to express the business or domain value of satisfying the rule $\phi_i$.

To use this gain formulation, we are faced with two practical challenges: (1) we do not know the probabilities $p_j$ for Eq. 1, since we do not know the correctness of the update $r_j$ beforehand, and (2) we do not know the desired clean database $D^{opt}$ for computing Eq. 2, since that is the goal of the cleaning process in the first place.

**User Model**: To approximate $p_j$, we learn and model the user as we obtain his/her feedback for the suggested updates. $p_j$ is approximated by the prediction probability, $\tilde{p}_j$, of having $r_j$ correct (learning user feedback is discussed in Section 4.2). Since initially there is no feedback, we assign $s_j$ to $\tilde{p}_j$, where $s_j \in [0, 1]$ is a score that represents the repairing algorithm certainty about the suggested update $r_j$.

**Estimating Update Benefit**: To compute the overall quality loss $L$ in Eq. 3, we need to first compute the quality loss with respect to a particular rule $\phi$, i.e., $ql(D|c, \phi)$ in Eq. 2. To this end, we approximate the numerator and denominator separately. The numerator expression, which represents the difference between the numbers of tuples satisfying $\phi$ in $D^{opt}$ and $D$, respectively, is approximated using $D$'s violations with respect to $\phi$. Thus, we use the expression $vio(D, \{\phi\})$ (cf. Definition 1) as the numerator in Eq. 2.

The main approximation we made is to assume that the updates within a group $c$ are independent. Hence to approximate the denominator of Eq. 2, we assume further that there is only one suggested update $r_j$ in $c$. The effect of this last assumption is that we consider two possible clean desired databases—one in which $r_j$ is correct, denoted by $D^{r_j}$, and another one in which $r_j$ is incorrect, denoted by $D^{\bar{r_j}}$. Consequently, there are two possibilities for the denominator of Eq. 2, each with a respective probability $p_j$ and $(1 - p_j)$. Our evaluations show that despite our approximations, our approach produces a good ranking of the groups of updates.

We apply this approximation independently for each $r_j \in c$ and estimate the quality loss $ql$ as follows:

$$E[ql(D|c, \phi)] = \sum_{r_j \in c} [\tilde{p}_j \cdot \frac{vio(D, \{\phi\})}{|D^{r_j} \models \phi|} + (1 - \tilde{p}_j) \frac{vio(D, \{\phi\})}{|D^{\bar{r_j}} \models \phi|}] \quad (4)$$

where we approximate $p_j$ with $\tilde{p}_j$.

The expected loss in data quality for the database $D$, given the suggested group of updates $c$, can be then approximated based on Eq. 3 by replacing $ql$ with $E[ql]$ obtained from Eq. 4:

$$E[L(D|c)] = \sum_{\phi_i \in \Sigma} w_i \sum_{r_j \in c} \left[ \tilde{p}_j \frac{vio(D, \{\phi\})}{|D^{r_j} \models \phi|} + (1 - \tilde{p}_j) \frac{vio(D, \{\phi\})}{|D^{\bar{r_j}} \models \phi|} \right] \quad (5)$$

We can also compute the expected loss for $D^{r_j}$ and $D^{\bar{r_j}}$ using Eq. 3 and Eq. 5 as follows: $E[L(D^{r_j})] = \sum_{\phi_i \in \Sigma} w_i \cdot \frac{vio(D^{r_j}, \{\phi_i\})}{|D^{r_j} \models \phi_i|}$ where we use $\tilde{p}_j = 1$ since in $D^{r_j}$ we know that $r_j$ is correct and $E[L(D^{\bar{r_j}})] = \sum_{\phi_i \in \Sigma} w_i \cdot \frac{vio(D^{\bar{r_j}}, \{\phi_i\})}{|D^{\bar{r_j}} \models \phi_i|}$ where we use $\tilde{p}_j = 0$ since in $D^{\bar{r_j}}$ we know that $r_j$ is incorrect.

Finally, using Eq. 1 and substituting $L(D|c)$ with $E[L(D|c)]$ from Eq. 5, we compute an estimate for the data quality gain of acquiring feedback for the group $c$ as follows:

$$E[g(c)] = E[L(D|c)] - \sum_{r_j \in c} [\tilde{p}_j \; E[L(D^{r_j})] + (1-\tilde{p}_j)E[L(D^{\bar{r_j}})]]$$

$$= \sum_{\phi_i \in \Sigma} w_i \sum_{r_j \in c} \left[\tilde{p}_j \frac{vio(D, \{\phi_i\})}{|D^{r_j} \models \phi_i|} + (1-\tilde{p}_j)\frac{vio(D, \{\phi_i\})}{|D^{\bar{r_j}} \models \phi_i|}\right]$$

$$- \sum_{r_j \in c}$$

$$\left[\tilde{p}_j \sum_{\phi_i \in \Sigma} w_i \frac{vio(D^{r_j}, \{\phi_i\})}{|D^{r_j} \models \phi_i|} + (1-\tilde{p}_j) \sum_{\phi_i \in \Sigma} w_i \frac{vio(D^{\bar{r_j}}, \{\phi_i\})}{|D^{\bar{r_j}} \models \phi_i|}\right]$$

Note that $vio(D, \{\phi_i\}) - vio(D^{\bar{r_j}}, \{\phi_i\}) = 0$ since $D^{\bar{r_j}}$ is the database resulting from rejecting the suggested update $r_j$ which will not modify the database. Therefore, $D^{\bar{r_j}}$ is the same as $D$ with the same violations. After a simple rearrangement, we obtain the final formula to compute the estimated gain for $c$:

$$E[g(c)] = \sum_{\phi_i \in \Sigma} \left[w_i \sum_{r_j \in c} \tilde{p}_j \frac{vio(D, \{\phi_i\}) - vio(D^{r_j}, \{\phi_i\})}{|D^{r_j} \models \phi_i|}\right] \quad (6)$$

The final formula in Eq. 6 is intuitive by itself and can be justified by the following. The main objective to improve the quality is to reduce the number of violations in the database. Therefore, the difference in the amount of database violations as defined in Definition 1, before and after applying $r_j$, is a major component to compute the update benefit. This component is computed, under the first summation, for every rule $\phi_i$ as a fraction of the number of tuples that would be satisfying $\phi_i$, if $r_j$ is applied. Since the correctness of the repair $r_j$ is unkown, we cannot use the term under the first summation as a final benefit score. Instead, we compute the expected update benefit by approximating our *certainty* about the benefit by the prediction probability $\tilde{p}_j$.

**Example:** For the example in Figure 1, assume that the repairing algorithm generated 3 updates to replace the value of the CT attribute by 'Michigan City' in $t_2, t_3$ and $t_4$. Assume also that the probabilities, $\tilde{p}_j$, for each of them are 0.9, 0.6, and 0.6, respectively. The weights $w_i$ for each $\phi_i$, $i = 1, \ldots, 5$ are $\{\frac{4}{8}, \frac{1}{8}, \frac{2}{8}, \frac{1}{8}, \frac{3}{8}\}$. Due to this modifications only $\phi_1$ will have their violations affected. Then for this group of updates, the estimated benefit can be computed as follow using Eq. 6: $\frac{4}{8} \times (0.9 \times \frac{4-3}{1} + 0.6 \times \frac{4-3}{1} + 0.6 \times \frac{4-3}{1}) = 1.05$. □

### 4.2 Active Learning Ordering

One way to reduce the cost of acquiring user feedback for verifying each update is to relegate the task of providing feedback to a machine learning algorithm. The use of a learning component in GDR is motivated by the existence of correlations between the original data and the correct updates. If these correlations can be identified and represented in a classification model, then the model can be trained to predict the correctness of a suggested update and hence replace the user for similar (future) situations.

As stated earlier, GDR provides groups of updates to the user for feedback. Here, we discuss how the updates within a group will be ordered and displayed to the user, such that user feedback for the top updates would strengthen the learning component's capability to replace the user for predicting the correctness for the rest of the updates.

**Interactive Active Learning Session:** After ranking the groups of updates, the user will pick a group $c$ that has a high score $E[g(c)]$. The learner orders these updates such that those that would most benefit, i.e., improve the model prediction accuracy, from labeling come first. The updates are displayed to the user along with their learner predictions for the correctness of the update. The user will then give feedback on the top $n_s$ updates, that she is sure about, and inherently correct any mistakes made by the learner. The newly labeled examples in $n_s$ are added to the learner training dataset $T_r$ and the active learner is retrained. The learner then provides new predictions and reorder the currently displayed updates based on the training examples obtained so far. If the user is not satisfied with the learner predictions, the user will then give feedback on another $n_s$ updates from $c$. This interactive process continues until the user is either satisfied with the learner predictions, and thus delegates the remaining decisions on the suggested updates in $c$ to the learned model, or the updates within $c$ are all labeled, i.e., verified, by the user.

**Active Learning:** In the learning component, there is a machine learning algorithm that constructs a classification model. Ideally, we would like to learn a model to automatically identify correct updates without user intervention. *Active learning* is an approach to learning models in such situations where unlabeled examples (i.e. suggested updates) is plentiful but there is a cost to labeling examples (acquiring user feedback) for training.

By delegating some decisions on suggested updates to the learned models, GDR is allowing for "automatic" repairing. However, there is a guarantee to correctly repair the data that is inherently provided by the active learning process to learn accurate classifiers to predict the correctness of the updates. The user is the one to decide whether the classifiers are accurate while inspecting the suggestions.

**Learning User Feedback:** The learning component predicts for a suggested update $r = \langle t, A, v, s \rangle$ one of the following predictions, which corresponds to the expected user feedback. (i) *confirm*, the value of $t[A]$ should be $v$. (ii) *reject*, $v$ is not a valid value for $t[A]$ and GDR needs to find another update. (iii) *retain*, $t[A]$ is a correct value and there is no need to generate more updates for it. The user may also suggest new value $v'$ for $t[A]$ and GDR will consider it as a confirm feedback for the repair $r' = \langle t, A, v', 1 \rangle$.

In the learning component, we learn a set of classification models $\{M_{A_1}, \ldots, M_{A_n}\}$, one for each attribute $A_i \in attr(R)$. Given a suggested update for $t[A_i]$, model $M_{A_i}$ is consulted to predict user feedback. The models are trained by examples acquired incrementally from the user. We present here our choices for data representation (input to the classifier), classification model, and *learning benefit* scores.

**Data Representation:** For a given update $r = \langle t, A_i, v, s \rangle$ and user feedback $\mathcal{F} \in \{\text{confirm, reject, retain}\}$, we construct a training example for model $M_{A_i}$ in the form $\langle t[A_1], \ldots, t[A_n], v, \mathcal{R}(t[A_i], v), \mathcal{F}\rangle$. Here, $t[A_1], \ldots, t[A_n]$ are the original attributes' values of tuple $t$ and $\mathcal{R}(t[A_i], v)$[1] is a function that quantifies the relationship between $t[A_i]$ and its suggested value $v$.

Including the original dirty tuple along with the suggested update value enables the classifier to model associations between original attribute values and suggested values. Including the relationship function, $\mathcal{R}$, enables the classifier to model associations based on similarities that do not depend solely on the values in the original database instance and the suggested updates.

**Active Learning Using Model Uncertainty:** Active learning starts with a preliminary classifier learned from a small set of labeled training examples. The classifier is applied to the unlabeled examples and a scoring mechanism is used to estimate the most valuable example to label next and add to the training set. Many criteria have been proposed to determine the most valuable examples for labeling (e.g, [20, 23]) by focusing on selecting the examples whose predictions have the largest *uncertainty*.

---
[1]We use a string similarity function.

One way to derive the uncertainty of an example is by measuring the disagreement amongst the predictions it gets from a committee of $k$ classifiers [19]. The committee is built so that the $k$ classifiers are slightly different from each other, yet they all have similar accuracy on the training data. For an update $r$ to be classified by label $\mathcal{F} \in \{\text{confirm, reject, retain}\}$, it would get the same prediction $\mathcal{F}$ from all members. The uncertain ones will get different labels from the committee and by adding them in the training set the disagreement amongst the members will be lowered.

In our implementation, each model $M_{A_i}$ is a random forest which is an *ensemble* of decision trees [5] that are built in a similar way to construct a committee of classifiers. Random forest learns a set of $k$ decision trees. Let the number of instances in the training set be $N$ and the number of attributes in the examples be $M$. Each of the $k$ trees are learned as follows: randomly sample *with replacement* a set $S$ of size $N' < N$ from the original data, then learn a decision tree with the set $S$. The random forest algorithm uses a standard decision-tree learning algorithm with the exception that at each attribute split, the algorithm selects the best attribute from a random subsample of $M' < M$ attributes. We used the WEKA[2] random forest implementation with $k = 10$ and default values for $N'$ and $M'$.

**Computing Learning Benefit Score:** To classify an update $r = \langle t, A_i, v, s \rangle$ with the learned random forest $M_{A_i}$, each tree in the ensemble is applied separately to obtain the predictions $\mathcal{F}_1, \ldots, \mathcal{F}_k$ for $r$, then the majority prediction from the set of trees is used as the output classification for $r$. The learning benefit or the uncertainty of predictions of a committee can be quantified by the entropy on the fraction of committee members that predicted each of the class labels.

**Example.** Assume that $r_1, r_2$ are two candidate updates to change the CT attribute to 'Michigan City' in tuples $t_2, t_3$. The model of the CT attribute, $M_{\text{CT}}$, is a random forest with $k = 5$. By consulting the forest $M_{\text{CT}}$, we obtain for $r_1$, the predictions {confirm, confirm, confirm, reject, retain}, and for $r_2$, the predictions {confirm, reject, reject, reject, reject}. In this case, the final prediction for $r_1$ is 'confirm' with an uncertainty score of 0.86 ($= -\frac{3}{5} \times \log_3 \frac{3}{5} - \frac{1}{5} \times \log_3 \frac{1}{5} - \frac{1}{5} \times \log_3 \frac{1}{5}$) and for $r_2$ the final prediction is 'reject' with an uncertainty score of 0.45. In this case, $r_1$ will appear to the user before $r_2$ because it has higher uncertainty. □

## 5. EXPERIMENTS

In this section, we present a thorough evaluation of the GDR framework, which has already been demonstrated in [22]. Specifically, we show that the proposed ranking mechanism converges quickly to a better data quality state. Moreover, we assess the trade-off between the user efforts and the resulting data quality in Appendix B.

We used in the experiments two real-world datasets referred to as Dataset 1 and 2, each with about 20,000 records. Appendix B provides details on the datasets, ground truth and the quality rules.

**User interaction simulation.** We simulated user feedback to suggested updates by providing answers as determined by the ground truth.

**Data quality state metric.** We report the improvement in data quality through computing the loss (Eq. 3). We consider the ground truth as the desired clean database $D^{opt}$.

---
[2]http://www.cs.waikato.ac.nz/ml/weka/

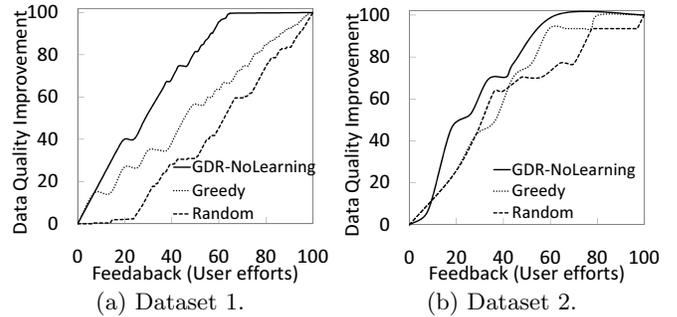

(a) Dataset 1.    (b) Dataset 2.

**Figure 3: Comparing VOI-based ranking in GDR (*GDR-NoLearning*) to other strategies against the amount of feedback. Feedback is reported as the percentage of the maximum number of verified updates required by an approach. Our application of the VOI concept shows superior performance compared to other naïve ranking strategies.**

### 5.1 VOI Ranking Evaluation

The objective here is to evaluate the effectiveness and quality of the VOI-based ranking mechanism described in Section 4.1. In this experiment, we did not use the learning component to replace the user; the user will need to evaluate each suggested update. Recall that the grouping provides the user with related tuples and their corresponding updates that could help in a quick batch inspection by the user.

We compare in this experiment the following techniques:

- *GDR-NoLearning*: The GDR framework of Figure 2(a) without the learning component.
- *Greedy*: Here, we rank the groups according to their sizes. The rationale behind this strategy is that groups that cover larger numbers of updates may have high impact on the quality if most of the suggestions within them are correct.
- *Random*: The naïve strategy where we randomly order the groups; all update groups are equally important.

In Figure 3, we show the progress in improving the quality against the number of verified updates (i.e., the amount of feedback). The feedback is reported as a percentage of the total number of suggested updates through the interaction process to reach the desired clean database.

The ultimate objective of GDR is to minimize user effort while reaching better quality quickly. In Figure 3, the slope of the curves in the first of iterations with the user is the key component to the curve: the steeper the curve the better the ranking. As illustrated for both datasets, the *GDR-NoLearning* approach performs well compared to the *Greedy* and *Random* approaches. This is because the *GDR-NoLearning* approach perfectly identifies the most beneficial groups that are more likely to have correct updates. While the *Greedy* approach improves the quality, most of the content of the groups is sometimes incorrect updates leading to wasted user efforts. The *Random* approach showed the worst performance in Dataset 1, while for Dataset 2, it was comparable with the *Greedy* approach especially in the beginning of the curves. This is because in Dataset 2, most of the sizes of the groups were close to each others making the *Random* and *Greedy* approaches behave almost identically, while in Dataset 1 the groups sizes varies widely making the random choices ineffective. Finally, we notice that *GDR-NoLearning* is much better for Dataset 1 than for Dataset 2, because of two reasons related to the nature of the Dataset 2: (i) most of the initially suggested updates for Dataset 2 are correct, and (ii) the sizes of the groups in Dataset 2 are close to each other. The consequence is that any ranking strategy for Dataset 2 will not be far from the optimal.

The results reported above justify clearly the importance and effectiveness of the GDR ranking component. The *GDR-NoLearning* approach is well suited for repairing "very" critical data, where every suggested update has to be verified before applying it to the database.

## 5.2 GDR Overall Evaluation

Here, we evaluate GDR's performance when using the learning component to reduce user efforts. More precisely, we evaluate the VOI-based ranking when combined with the active learning ordering. For this experiment, we evaluate the following approaches:

- *GDR*: is the approach proposed in this paper. In each interactive session, the user provides feedback for the top ranked updates. The required amount of feedback per group is inversely proportional to the benefit score of the group (Eq. 6)—the higher the benefit the less effort from the user is needed, since most likely the updates are correct and there are very few uncertain updates for the learned model that would require user involvement. As such, we require that the user verifies $d_i$ updates for a group $c_i$, $d_i = E \times \left(1 - \frac{g(c_i)}{g_{max}}\right)$, where $E$ is the initial number of dirty tuples and $g_{max} = \max_{\forall c_j}\{g(c_j)\}$.
- *GDR-S-Learning*: Here, we eliminate the active learning from the system—the updates are grouped and then ranked using VOI-based scoring alone. User is solicited for a random selection of updates within each group, instead of being ordered by uncertainty. However, all of the user feedback is used to train the learning component, which then replaces the user on deciding for the remaining updates in the group. *GDR-S-Learning* is included to assess the benefit of the active learning aspect of our framework, compared with traditional passive learning.
- *Active-Learning*: In this approach, we eliminate the grouping and their ranking from the GDR framework. In other words, we neither group the updates nor use VOI-based scores for ranking. We only solicit user feedback for updates ordered with the learner uncertainty scores. The user is required to provide feedback for the top update and then the learning component is updated to reorder the updates for the user in an iterative fashion. The resulting learned model is applied for predicting the remaining suggested updates and the database is updated accordingly. We report the quality improvement for different amount of feedbacks. This approach is included to assess the benefit of the grouping and the VOI-based ranking mechanisms compared with using only an active learning approach.
- *GDR-NoLearning*: This approach is the one described in the previous experiment; It provides a baseline to assess the utility of machine learning aspect for GDR.
- *Automatic-Heuristic*: The BatchRepair method described in [7] for automatic data repair using CFDs.

In Figure 4, we report the improvement in data quality as the amount of feedback increases. Assuming that the user can afford verifying at most a number of updates equal to the number of initially identified dirty tuples (6000 for Dataset 1 and 3000 for Dataset 2), we report the amount of feedbacks as a percentage of this number. The results show that *GDR* achieves superior performance compared with the other approaches; For Dataset 1, GDR gain about 90% improvement with 20% efforts or verifying about 1000 updates. For Dataset 2, about 94% quality improvement was gained with 30% efforts or verifying about 1000 updates.

In Dataset 1, *Active Learning* is comparable to *GDR* only in the beginning of the curve until reaching about 70% quality improvement. *GDR-S-Learning* starts to outperform *Active Learning* after about 45% user effort. The *Heuristic* approach repairs the database without user feedback, therefore, it produces a constant result. Note that the quality improvement achieved by the *Heuristic* approach is attained by *GDR* with about 10% user effort, i.e., giving feedback for updates numbering about 10% of the initial set of dirty tuples in the database. The *GDR-NoLearning* approach does improve the quality of the database, but not as quickly as any of the approaches that use learning methods. In comparison to Figure 3, the final performance of *GDR-NoLearning* is 100%, assuming all required feedback were obtained. *GDR* involves learning which allows for automatic updates to be applied and hence opens the door for some mistakes to occur. Thus, the 100% accuracy may not be reached.

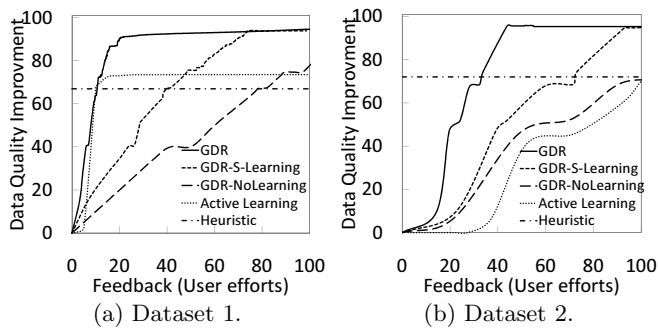

Figure 4: Overall evaluation of GDR compared with other techniques. The combination of the VOI-based ranking with the active learning was very successful in efficiently involving the user. The user feedback is reported as a percentage of the initial number of the identified dirty tuples.

For Dataset 2, similar results were achieved. However, the *Active Learning* approach was not as successful as for Dataset 1. This is due to the randomness nature of the errors in this dataset, which resulted in fewer correlations between these errors that could be learned by the model. Due to the wider array of real-world dependencies in Dataset 1, the machine learning methods were more successful and achieved better performance. For example, some hospitals located on the boundary between two zip codes have their zip attributes dirty; this is most likely due to a data entry confusion on where they are really located.

The superior performance of *GDR* is justified by the following: for a single group of updates, using the learner uncertainty to select updates can effectively strengthen the learned model predictions as these "uncertain" updates are more important for the model. In *GDR-S-Learning*, randomly inspecting updates from the groups provided by the VOI-based ranking does enhance the learned model. However, more user effort is wasted in verifying less important updates according to the learning benefit. For the *Active Learning* approach, it is apparent that having the user spend more effort does not help the learned model due to the model over fitting problem. This problem is avoided in both *GDR* and *GDR-S-Learning* approaches because of the grouping provided by the GDR framework. The grouping provides the learned model a mechanism to adapt locally to the current group, which in turn provides the necessary guidance for the model to *strongly* learn the associations for a highly beneficial group rather than just *weakly* learning the associations for a wide variety of cases. This is also the reason that the *GDR-S-Learning* eventually outperforms the *Active Learning* with an increase in user effort.

This experiment demonstrates the importance of the learning component for achieving a faster convergence to a better quality. The results support our initial hypothesis about the existence of correlations between the dirty and correct versions of the tuples in real-world data. Also, the combination of VOI-based ranking with active learning improves over the traditional active learning mechanism.

## 6. RELATED WORK

Existing data repair techniques have mostly focused on automatically repairing data by finding another database that is consistent and minimally different from the original database (e.g., [2, 7, 16]).

The updates generation in GDR is somehow similar to the repair approach of [7], which consults the user to either introduce a new CFD in the repairing algorithm or to manually update the data, and no machine learning is included to help in the cleaning process itself. GDR goes well beyond that by interactively generating and then grouping updates for the sake of efficiently involving the user.

The initial step to suggest updates for all dirty tuples is a time consuming process. Therefore, we discuss and evaluate in a workshop paper [21] a mechanism for ranking the rules such that smaller subsets of the dirty tuples are processed to find their suggested updates in each user interactive session.

A recent work to repair critical data with quality guarantee was introduced in [11]. While addressing a similar problem as GDR, the setting is different. In [11] it is assumed that a reference correct data exists and the user is required to specify certain attributes to be correct across the entire dataset. Moreover, the proposed solution relies on a pre-specified set of editing rules. This is not the case for GDR. GDR requires only a set of data quality rules.

Most existing systems for data cleaning provide tools for data exploration and transformation without taking advantage of recent efforts on automatic data repair. Usually, the repair actions are "explicitly specified by the user". For example, AJAX [12] proposes a declarative language to eliminate duplicates during data transformations. Potter's Wheel [17] combines data transformations with the detection of errors in the form of irregularities. None of these systems efficiently leverage user feedback, like GDR, by either ranking or using learning mechanisms, and moreover, the user does not necessarily have to explicitly specify updates.

Previous work on soliciting user feedback to improve data quality focuses on identifying correctly matched references in a large scale integrated data (e.g., [14, 19]). In [14], a decision theoretic framework similar to ours has been proposed to rank candidate reference matches to improve the quality of query responses in dataspaces, but it cannot be applied in a constrained repair framework for relational database. [19] introduced an active-learning based approach to build a generic matching function for identifying duplicate records.

Another closely related area is to solicit user feedback to improve the prediction quality of a learning model by taking into account data acquisition costs, for example selective supervision [15], which combines decision theory with active learning for the learned model benefit.

The uniqueness of our work resides in combining (i) the certainty of a update, which is derived from an automatic repairing algorithm using the repair evaluation function, and (ii) the uncertainty of a learner to accurately predict the correctness of a update, in a ranking mechanism that uniformly and judiciously balances between them with the goal of improving the data quality as quickly as possible.

## 7. CONCLUSION AND FUTURE WORK

We presented GDR, a framework that combines constraint-based repair techniques with user feedback through an interactive process. The main novelty of GDR is to solicit user feedback for the most useful updates using a novel decision-theoretic mechanism combined with active learning. The aim is to move the quality of the database to a better state as far as the data quality rules are concerned.

Our experiments show very promising results in moving the data quality forward with minimal user involvement.

Our future work includes extending GDR to support more types of data quality rules other than CFDs like CINDs [4], Matching dependencies [8], and Matching Rules [10]. More research challenges may emerge from supporting various kinds of rules and their interactions to produce good updates. Moreover, we are investigating approach for the guided discovery of the rules from dirty data.

# APPENDIX

## A. UPDATES GENERATION IMPLEMENTATION

### A.1 CFD: Overview

A CFD $\phi$ over $R$ can be represented by $\phi : (X \to Y, t_p)$, where $X$ and $Y \in attr(R)$, $X \to Y$ is a standard functional dependency (FD), referred to as FD *embedded* in $\phi$, and $t_p$ is a tuple *pattern* containing all attributes in $X$ and $Y$. For each $A \in (X \cup Y)$, the value of the attribute $A$ for the tuple pattern $t_p$, $t_p[A]$, is either a constant $'a' \in dom(A)$, or $'-'$ which represents a variable value. We denote $X$ as $LHS(\phi)$ (left hand side) and $Y$ as $RHS(\phi)$ (right hand side). Examples for CFD rules are provided in Figure 1.

To denote that a tuple $t \in D$ matches a particular pattern $t_p$, the symbol $\asymp$ is defined on data values and $'-'$. We write $t[X] \asymp t_p[X]$ iff for each $A \in X$, either $t[A] = t_p[A]$ or $t_p[A] = '-'$. For example, (Sherden RD, Fort Wayne, IN) $\asymp$ (−, Fort Wayne, −). We assume that CFDs are provided in the normal form [7], i.e., $\phi : (X \to A, t_p)$, $A \in attr(R)$ and $t_p$ is a single pattern tuple.

A CFD $\phi : (X \to A, t_p)$ is said to be *constant*, if $t_p[A] \neq '-'$. Otherwise, $\phi$ is a *variable* CFD. For example in Figure 1, $\phi_1$ is a constant CFD, while $\phi_5$ is a variable CFD.

A database instance $D$ *satisfies* the constant CFD $\phi = (X \to A, t_p)$, denoted by $D \models \phi$, iff for each tuple $t \in D$, if $t[X] \asymp t_p[X]$ then $t[A] = t_p[A]$. If $\phi$ is a variable CFD, then $D \models \phi$ iff for each pair of tuples $t_1, t_2 \in D$, if $t_1[X] = t_2[X] \asymp t_p[X]$ then $t_1[A] = t_2[A] \asymp t_p[A]$. This means that if $t_1[X]$ and $t_2[X]$ are equal and match the pattern $t_p[X]$, then $t_1[A]$ and $t_2[A]$ must also be equal to each other. CFDs address a single relation only. However, the repairing algorithm that uses CFDs is applicable to general relational schemas by simply repairing each relation in isolation.

### A.2 Resolving CFD Violations

A dirty tuple $t$ may violate a CFD $\phi = (R : X \to A, t_p)$ in $\Sigma$ following two possible cases [7]:

- Case 1: $\phi$ is a constant CFD (i.e., $t_p[A] = a$, where $a$ is a constant) and $t[X] \asymp t_p[X]$ but $t[A] \neq a$.
- Case 2: $\phi$ is a variable CFD, $t[X] \asymp t_p[X]$, and $\exists t'$ such that $t'[X] = t[X] \asymp t_p[X]$ but $t[A] \neq t'[A]$.

The latter case is similar to the violation of a standard FD. Accordingly, given a set $\Sigma$ of CFDs, the dirty tuples can be immediately identified and stored in the *DirtyTuples* list.

To resolve a violation of a CFD $\phi = (R : X \to A, t_p)$ by a tuple $t$, we proceed as follows: For case 1, we either modify the $RHS(\phi)$ attribute such that $t[A] = t_p[A]$ or we change some of the attributes in $LHS(\phi)$ such that $t[X] \not\asymp t_p[X]$. For case 2, we either modify $t[A]$ (resp. $t'[A]$) such that $t[A] = t'[A]$ or we change some $LHS(\phi)$ attributes $t[X]$ (resp. $t'[X]$) such that $t[X] \neq t'[X]$ or $t[X] \not\asymp t_p[X]$ (resp. $t'[X] \not\asymp t_p[X]$).

**Example:** In Figure 1, the normal form of $\phi_1 :$ (ZIP $\to$ CT, STT, $\{46360 \parallel MichiganCity, IN\}$) would be $\phi_{1,1} :$ (ZIP $\to$ CT, $\{46360 \parallel MichiganCity\}$) and $\phi_{1,2} :$ (ZIP $\to$ STT, $\{46360 \parallel IN\}$).

$t_2$ violates $\phi_{1,1} :$ (ZIP $\to$ CT, $\{46360 \parallel MichiganCity\}$) following case 1. Thus, a suggested update by changing $RHS(\phi_{1,1})$ is to replace 'Westville' by 'Michigan City' in $t_2$[CT], while another update by changing LHS($\phi_{1,1}$) is to replace '46360' by '46391' in $t_2$[ZIP], for example. $t_5, t_6$ both violate $\phi_5$ following case 2. A possible update is to change RHS($\phi_5$) by modifying $t_5$[ZIP] to be '46825' instead of '46391'. Yet, another possible update is to make a change in LHS($\phi_5$). For example, by changing $t_5$[STR] or $t_5$[CT] to another value. □

### A.3 Update Evaluation Function

Since there may be many possible ways to clean a dirty tuple, we need an evaluation function to select the "best" updates. We follow the same evaluation approach used in [2] and [7]. Given an update $r$ to modify $t[A] = v$ such that $t[A] = v'$, we compute the update evaluation score $s$ as the similarity between $v$ and $v'$. This can be done based on the edit distance function $dist_A(v, v')$ as follows

$$s(r) = sim(v, v') = 1 - \frac{dist_A(v, v')}{\max(|v|, |v'|)}. \quad (7)$$

where $|v|, |v'|$ denote the size of $v, v'$, respectively. The intuition here is that, the more accurate $v'$, the more it is close to $v$. $s(r)$ is in the range $[0..1]$ and any domain specific similarity function can be used for this purpose. Finally, the update can be composed in the tuple form $r = \langle t, A, v', s(r) \rangle$.

### A.4 Generating Updates

We now show how to use CFDs to generate updates for each potentially dirty attribute $B$ in $t \in DirtyTuples$. The generated updates are tuples in the form $\langle t, B, v, s \rangle$, where $v$ is the suggested repair value for $t[B]$ and $s$ is the repair evaluation score from Eq. 7.

The suggested updates correspond to attribute value modifications, which are enough for CFDs violations [7]. For each dirty tuple $t$, we store the list of violated rules in $t.vioRuleList$. Furthermore, for each pair $\langle t, B \rangle$, we keep a list of values $\langle t, B \rangle.preventedList$, which contains values for $t[B]$ that are confirmed as wrong. Thus, when searching a new suggestion for $t[B]$, the values in $\langle t, B \rangle.preventedList$ are discarded. Also, we keep a flag $\langle t, B \rangle.Changeable$ that is set to False when the value in $t[B]$ was confirmed to be correct.

Initially, we assume that each attribute value is incorrect for all $t \in DirtyTuples$ and proceed by searching for the best update value that provides the best score according to Eq. 7. This can be performed by calling Algorithm 1, $UpdateAttributeTuple(t, B)$ for all $t \in DirtyTuples$ and $B \in attr(R)$.

$UpdateAttributeTuple$ described in Algorithm 1 finds the best update value for $t[B]$ by exploring three possible scenarios:

1. $B = A$ for some violated CFD $\phi = (X \to A, t_p)$ and $t_p[A] \neq '-'$ (i.e., $\phi$ is a constant CFD): This corresponds to case 1 of rule violations where $t[X] \asymp t_p[X]$ and $t[A] \not\asymp t_p[A]$. In this scenario, a value $v = a$ is suggested (lines 4-6).

2. $B = A$ for some violated CFD $\phi = (X \to A, t_p)$ and $t_p[A] = '-'$ (i.e., $\phi$ is a variable CFD): This corresponds to case 2 of rule violations where $t[X] \asymp t_p[X]$ and $t[A] \asymp t_p[A]$ and there exists another tuple $t'$ that violates $\phi$ with $t$, i.e., $t'[X] \asymp t[X]$ but $t'[A] \not\asymp t[A]$. In this scenario, a value $v = t'[A]$ is suggested (lines 7-9).

3. $B \in LHS(\phi)$ for some violated CFD $\phi = (X \to A, t_p)$: This corresponds to either case 1 or case 2 of rule violations. In this scenario, we look for a value $v$ that maximizes the repair evaluation score $sim(t[B], v)$ (Eq. 7.) The aim is to select semantically related values by first using the values in the CFDs, then searching in the tuples identified by the pattern $t[X \cup A - \{B\}]$ (lines 11-13).

In each of the above scenarios, the value $v \notin \langle t, B \rangle.preventedList$. Finally, a repair tuple is composed $\langle t, B, v, s \rangle$ and inserted into $PossibleUpdates$ in line 14

**Example:** In Figure 1, $t_5$ violates $\phi_4$ and when repairing the attribute CT $\in RHS(\phi_4)$, a suggested update according to Scenario 1 will be 'Westville'. Also, $t_5$ violates $\phi_5$ and when repairing the attribute ZIP $\in RHS(\phi_5)$, a suggested update will be 46825 according to Scenario 2. When repairing the attribute STR $\in LHS(\phi_5)$, a suggested value from the domain $dom(\text{STR})$ can be 'Sherden RD' according to Scenario 3. □

**Analysis:** We assume that a tuple is violating $|t.vioTupleList|$ rules. From Definition 1, we know that if a tuple $t$ is violating a variable CFD $\phi$, then $vio(t, \phi)$ is the number of tuples that violate $\phi$ with $t$. Each of the scenarios in Algorithm 1 can be analyzed as follows: Scenario 1 requires $O(1)$ operations to suggest updates from each constant CFD violation. Scenario 2 requires $O(vio(t, \phi))$ to suggest a value from each set of tuples violating a variable rule $\phi$ with $t$. Scenario 3 requires searching the values in the rules and in the domain of attribute $A$, i.e., the worst case is $O(|dom(A)| + \Sigma)$. Then, the best case of running UpdateAttributeTuple is $O(|t.vioTupleList|)$ and the worst case is $O(|t.vioTupleList| \times (|dom(A)| + \Sigma))$.

---

**Algorithm 1** UpdateAttributeTuple (Tuple $t$, Attribute $B$)

---
1: **if** $\langle t, B \rangle.Changeable =$ false **then return**;
2: best_s = 0 ; $v = null$
3: **for all** $\phi = (X \rightarrow A, t_p) \in t.vioRuleList$ **do**
4:   **if** $B = A \wedge t_p[A] \neq$ '−' **then**
5:     cur_s = $sim(t[A], t_p[A])$     {scenario 1}
6:     **if** cur_s > best_s **then** { best_s = cur_s; $v = t_p[A]$ }
7:   **else if** $B = A \wedge t[A] =$ '−' **then**
8:     $\langle best\_s, v \rangle =$ getValueForRHS($\phi, A, t,$ best_s) {scenario 2}
9:   **end if**
10: **end for**
11: **if** $\exists \phi = (X \rightarrow A, t_p) \in t.vioRuleList$ s.t. $B \in X$ **then**
12:   $\langle best\_s, v \rangle =$ getValueForLHS($A, t,$ best_s) {scenario 3}
13: **end if**
14: **if** $v \neq null$ **then**
15:   $PossibleUpdates = PossibleUpdates \cup \{\langle t, B, v, s = sim(t[B], v) \rangle\}$
16: **end if**

---

## A.5 Updates Consistency Manager

Once a repair $r = \langle t, B, v, s \rangle$ is confirmed to be correct, either by the user or the learning component, it is applied immediately to the database to get a new database instance. Consequently, some of the already suggested updates may become inconsistent since they were generated according to a different database instance.

The consistency manager needs to maintain two invariants: (i) There is no tuple $t \in D$ such that $t \not\models \phi$ for any $\phi \in \Sigma$, and $t \notin DirtyTuples$. (ii) There is no update $r \in PossibleUpdates$ such that $r$ depends on data values that have been modified in the database. In the following, we provide the detailed steps of the consistency manager procedure that we implemented in GDR. Given an update $r = \langle t, B, v, s \rangle$ along with the feedback $\in$ {confirm, reject, retain}:

1. If the feedback is to retain the current value $t[B]$, then we set $\langle t, B \rangle.Changeable =$ false to stop looking for updates for $t[B]$.

2. If the feedback is to reject the update, i.e., $t[B]$ cannot be $v$, then $v$ is added immediately to the list $\langle t, B \rangle.PreventedList$. This is followed by a call to $UpdateAttributeTuple(t, B)$ to find another update for $t[B]$.

3. If the feedback confirms that $t[B]$ must be $v$, then the update is applied to the database immediately and we stop generating updates for $t[B]$ by setting $\langle t, B \rangle.Changeable =$ false. Afterward, we go through the rules that involve the attribute $B$ and update the necessary data structures to reflect the removed violations as well as new emerging violations. Particularly, for each $\phi : (X \rightarrow A, t_p) \in \Sigma$ where $B \in (X \cup A)$, we do the following:

   (a) If $t \not\models \phi$, then we consider two cases:
   
   i. $\phi$ is a constant CFD: If $\langle t, C \rangle.Changeable =$ false, $\forall C \in X$, i.e., all attributes in LHS($\phi$) have been confirmed as correct and are not changeable values, then RHS($\phi$) should be applied; we apply $t[A] = t_p[A]$ to the database directly, set $\langle t, A \rangle.Changeable =$ false, and remove $\phi$ from $t.vioRuleList$. If some of the LHS($\phi$) attribute values are changeable in $t$, then $\forall C \in (\{X \cup A\} / B)$ we add $\langle t, C \rangle$ to $RevisitList$. $\phi$ is added to $t.vioRuleList$, if it is not already there, and $t$ is added to the $DirtyTuples$ as well.
   
   ii. $\phi$ is a variable CFD: We add $\phi$ to $t.vioRuleList$ and then identify the tuples $t'$ that violate $\phi$ with $t$. Then for each $t'$, we add $\phi$ to $t'.vioRuleList$ and add $t'$ to the $DirtyTyples$. Also, we add $\langle t', C \rangle$ to the $RevisitList$, $\forall C \in \{X \cup A\}$ because this $\phi$ may be a new emerging violation for $t'$ and all the attributes are candidates to be wrong for $t'$.
   
   (b) If $t \models \phi$ while $\phi \in t.vioRuleList$, then $\phi$ originally was violated by $t$ before applying this update. Therefore, we remove $\phi$ from $t.vioRuleList$. If $\phi$ is a constant CFD, no further action is required. However, if $\phi$ is a variable CFD, we need to check the other tuples $t'$, which were involved with $t$ in violating $\phi$, and eventually update their $vioRuleList$. We remove $\phi$ from $t'.vioRuleList$ as long as $\not\exists t''$ s.t. $t', t'' \not\models \phi$, i.e., $t'$ is not involved in violating $\phi$ with another tuple $t''$.

4. Remove update $r = \langle t, C, v, s \rangle$ from the $PossibleUpdates$, if $\langle t, C \rangle \in RevisitedList$ or $\langle t, C \rangle.Changeable =$ false.

5. For every element $\langle t, C \rangle \in RevisitedList$, we call $UpdateAttributeTuple(t, C)$ to find another repair for $t[C]$.

6. Remove $t$ from $DirtyTuples$, if $t.vioRuleList$ is empty.

Note that the first update consistency invariant is maintained because of the following: A tuple $t$ may become dirty if it is modified or another tuple $t'$ is modified so that $t, t'$ violates some variable CFD $\phi \in \Sigma$. For a tuple $t$ and a CFD rule $\phi$, assuming that due to a database update $t \not\models \phi$, then $t$ must be in $DirtyTuples$ after applying Step 3a.

If $\phi$ is a constant CFD, then step 3(a)i should have been applied. If $t$ continues to violate $\phi$ it should be in $DirtyTuples$. If $\phi$ is a variable CFD, then step 3(a)ii should have been applied. There are two cases to consider: First, if $t$ is the tuple being repaired and $t \not\models \phi$, then it is added to $DirtyTuples$, if not already there. Second, if $t \not\models \phi$ because another tuple $t'$ was repaired (or modified), then step 3(a)ii should have been applied on $t'$. Thus all tuples involved with $t'$ in violating $\phi$, including $t$ will be added to *DirtyTuples*. Following the same rationale, step 3b maintains that $t.vioRuleList$ contains only rules that are being violated by $t$. Thus, Step 6 guarantees that the content of *DirtyTuples* corresponds to tuples involved in rules violation.

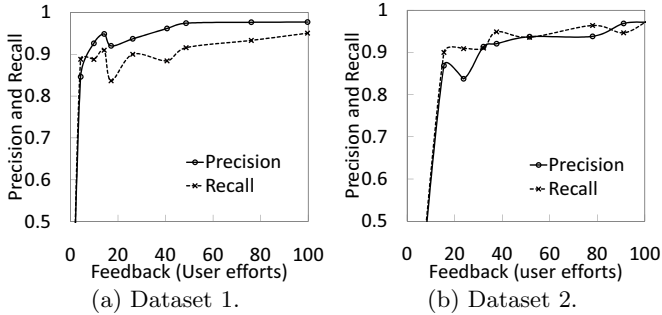

(a) Dataset 1.   (b) Dataset 2.

**Figure 5: Accuracy vs. user efforts.** As the user spends more effort with GDR, the overall accuracy is improved. The user feedback is reported as a percentage of the initial number of the identified dirty tuples.

The second update consistency invariant is maintained as well because of Steps 3(a)i, 4, and 5. These steps maintain a local list, *RevisitedList*, to hold tuple-attribute pairs, where their generated updates may depend on the applied update. In Step 3(a)i, changing the value of $t[B]$ may affect the update choice for the other attributes of $\phi$. For a variable CFD, Step 3(a)ii, all the tuples involved in the violations due to the modified value will need their attributes values to be revisited to find updates. Step 4 removes the corresponding updates from the *PossibleUpdates* and we proceed in Step 5 to get potentially new updates.

Note that Step 3 loops on the set of rules for the particular tuple $t$ that was updated. In Steps 3(a) and 3(b), we consider the immediate dependencies (consequences) of updating $t$ with respect to a single rule $\phi$. Particularly in Step 3(a), we check for new violations for $\phi$ that involve $t$, because it is the only change to the database. In Step 3(b), we check for already resolved violations for $\phi$ due to updating $t$. This local process to tuple $t$ that considers only a single rule $\phi$ at a time guarantees that the consistency manager will terminate and will not get into an infinite loop.

Since GDR is meant for repairing online databases, the consistency manager will need to be informed (e.g., through database triggers) with any new added or modified tuples. Every new tuple or modified values can be considered as an update and the above steps will proceed naturally.

## B. EXPERIMENTS SETTINGS

**Datasets.** In our experiments, we used two datasets, denoted as Dataset 1 and 2 respectively. Dataset 1 is a real world dataset obtained by integrating (anonymized) emergency room visits from 74 hospitals. Such patient data is used to monitor naturally occurring disease outbreaks, biological attacks, and chemical attacks. Since such data is coming from several sources, a myriad of data quality issues arise due to the different information systems used by these hospitals and the different data entry operators responsible for entering this data. For our experiments, we selected a subset of the available patient attributes, namely Patient ID, Age, Sex, Classification, Complaint, HospitalName, StreetAddress, City, Zip, State, and VisitDate. For Dataset 2, we used the adult dataset from the UCI repository (http://archive.ics.uci.edu/ml/). For our experiments, we used the attributes education, hours_per_week, income, marital_status, native_country, occupation, race, relationship, sex, and workclass.

**Ground truth.** To evaluate our technique against a ground-truth, we manually repaired 20,000 patient records in Dataset 1. We used address and zip code lookup web sites for this purpose. We assumed that Dataset 2, which is about 23,000 records, is already clean and hence can be used as our ground truth. We synthetically introduced errors in the attribute values as follows. We randomly picked a set of tuples, and then for each tuple, we randomly picked a subset of the attributes to perturb by either changing characters or replacing the attribute value with another value from the domain attribute values. All experiments are reported when 30% of the tuples are dirty.

**Data Quality Rules.** For Dataset 1, we used CFDs similar to what was illustrated in Figure 1. The rules were identified while manually repairing the tuples. For Dataset 2, we implemented the technique described in [9] to discover CFDs and we used a support threshold of 5%.

**Settings.** All the experiments were conducted on a server with a 3 GHz processor and 32 GB RAM running on Linux. We used Java to implement the proposed techniques and MySQL to store and query the records.

### B.1 Additional Experiments: User Efforts vs. Repair Accuracy

We evaluate GDR's ability to provide a trade-off between user effort and accurate updates. We use the precision and recall, where precision is defined as the ratio of the number of values that have been correctly updated to the total number of values that were updated, while recall is defined as the ratio of the number of values that have been correctly updated to the number of incorrect values in the entire database. Since we know the correct data, we can compute these values.

The user in this experiment affords only verifying $F$ updates, then GDR decide about the rest of the updates automatically. GDR asks the user to verify $d_i$ of the suggested updates in a group of repairs $c_i$, until we reach $F$.

In Figure 5, we report the precision and recall values resulting from repairing the database as we increase $F$ (reported as % of dirty tuples). For both datasets the precision and recall generally improve as $F$ increases. However, for Dataset 1, the precision is always higher than for Dataset 2. This is due to the lower accuracy of the learning component for Dataset 2, which stems from the random nature of the errors in Dataset 2. Overall, these results illustrate the benefit of user feedback—as the user effort increases, the repair accuracy increases.